\documentclass[12pt]{article}
\usepackage{amssym}

\font\elevenof=msbm10 at 11pt

\def\Z{\mbox{$\Bbb Z$}}
\def\I{\mbox{$\Bbb I$}}
\def\AG{${\cal A}(G(N))$}
\def\ap{a^{\dagger}}
\def\Qp{Q^{\dagger}}
\def\mod{\mathop{\rm mod}\nolimits}
\def\diag{\mathop{\rm diag}\nolimits}
\def \Al{${\cal A}^{(\lambda)}(G(N))$}

\title{
REDUCIBILITY AND BOSONIZATION OF PARASUPERSYMMETRIC AND
ORTHOSUPERSYMMETRIC QUANTUM MECHANICS}
\author{C. QUESNE\thanks{Directeur de recherches FNRS} \ AND N. VANSTEENKISTE \\
{\small \sl Physique Nucl\'eaire Th\'eorique et Physique Math\'ematique,}\\ {\small \sl
Universit\'e Libre de Bruxelles, Campus de la Plaine CP229,} \\ {\small \sl  Boulevard~du
Triomphe, B-1050 Brussels, Belgium} \\
{\small \sl E-mail: cquesne@ulb.ac.be}}
\date{ }
\begin{document}
\baselineskip=22pt plus 1pt minus 1pt
\maketitle

\begin{abstract} 
Order-$p$ parasupersymmetric and orthosupersymmetric quantum mechanics are
shown to be fully reducible when they are realized in terms of the generators of a
generalized deformed oscillator algebra and a $\mbox{\elevenof Z}_{p+1}$-grading
structure is imposed on the Fock space. The irreducible components provide $p+1$ sets
of bosonized operators corresponding to both unbroken and broken cases. Such a
bosonization is minimal.
\end{abstract}

\noindent
Running head: Reducibility and Bosonization

\noindent
PACS: 03.65.Fd, 11.30.Pb

\noindent
Keywords: quantum mechanics, supersymmetry, generalized deformed oscillator
\newpage
%
%
\section{Introduction}

Since the introduction of supersymmetric quantum mechanics by Witten~\cite{witten},
there has been a lot of interest in this field and its generalizations~\cite{cooper}. Among
the latter, one may quote parasupersymmetric (PSSQM)~\cite{rubakov, beckers90},
orthosupersymmetric (OSSQM)~\cite{khare}, pseudosupersymmetric
(PsSSQM)~\cite{beckers95}, and fractional supersymmetric quantum
mechanics~\cite{durand}. More recently, topological symmetries have also been
introduced~\cite{mosta}.\par
%
%
The standard way of realizing SSQM and its variants consists in combining bosons with
fermions or more exotic particles. Since fermionic operators can be realized in terms of
bosonic ones~\cite{naka}, it is straightforward to bosonize (i.e., to realize in terms of
only boson-like operators without fermion-like ones) $N=2$ SSQM from the well-known
Nicolai-Witten construction~\cite{witten, nicolai} in terms of two commuting sets of
independent operators. Some years ago, however, it was shown that SSQM can be
bosonized in a different way~\cite{plyu}. Contrary to the former procedure, the latter is
indeed a {\em minimal\/} bosonization in terms of a single bosonic degree of freedom.
Such a result was obtained from a realization of SSQM in terms of generalized
deformed bosonic oscillator operators~\cite{bonatsos} by imposing a
$\Z_2$-grading structure on the deformed bosonic oscillator Fock space, which can be
done, for instance, by restricting oneself to a Calogero-Vasiliev algebra~\cite{vasiliev}. In
such a framework, SSQM turns out to be fully reducible and its irreducible components
provide two sets of bosonized operators realizing either broken or unbroken SSQM.\par
%
%
An important open question is whether such results can be extended to SSQM variants.
As far as the bosonization is concerned, some partial results were recently obtained in
terms of generalized deformed oscillator algebras (GDOAs) ${\cal A}^{(\lambda)}(G(N))$
related to $C_{\lambda}$-extended oscillator ones, where $C_{\lambda} =
\Z_{\lambda}$ denotes the cyclic group of order $\lambda$ ($\lambda
 \in \{3, 4, 5, \ldots\}$)~\cite{cq98, cq00} (see also~\cite{daoud} for a related
work).\par
%
%
The aim of the present letter is to prove in full generality both the reducibility and the
resultant bosonization of order-$p$ PSSQM and OSSQM, where in the former case we
consider both the Rubakov-Spiridonov-Khare (RSK) approach~\cite{rubakov} and the
Beckers-Debergh (BD) one~\cite{beckers90}. For such a purpose, we shall
introduce some new matrix realizations of PSSQM and OSSQM in terms of GDOA
generators and impose a $\Z_{p+1}$-grading structure on the corresponding Fock space.
Restricting then ourselves to ${\cal A}^{(\lambda)}(G(N))$ with $\lambda = p+1$ will
provide us with some interesting examples of bosonized operators.\par
%
%
\section{\boldmath Generalized Deformed Oscillator Algebras}

There is an extensive literature on GDOAs to which we refer the reader (see
e.g.~\cite{katriel} and references quoted therein). Here we shall only review some of their
properties to be used in the next sections.\par
%
%
A GDOA may be defined as a nonlinear associative algebra \AG\ generated by the
operators $N = N^{\dagger}$, $\ap$, and $a = (\ap)^{\dagger}$, satisfying the
commutation relations
\begin{equation}
  [N, \ap] = \ap, \qquad [N, a] = - a, \qquad [a, \ap] = G(N),  \label{eq:GDOA}
\end{equation}
where $G(N) = [G(N)]^{\dagger}$ is some Hermitian function of $N$.\par
%
%
We restrict ourselves here to GDOAs possessing a bosonic Fock space representation. In
the latter, we may write $\ap a = F(N)$, $a \ap = F(N+1)$, where the structure function
$F(N) = [F(N)]^{\dagger}$ is such that
\begin{equation}
  G(N) = F(N+1) - F(N) \label{eq:G-F}
\end{equation}
and is assumed to satisfy the conditions
\begin{equation}
  F(0) = 0, \qquad F(n) > 0 \qquad \mbox{\rm if\ } n = 1, 2, 3, \ldots. 
  \label{eq:F-cond}  
\end{equation}
The carrier space $\cal F$ of such a representation can be constructed from a vacuum
state $|0\rangle$ (such that $a |0\rangle = N |0\rangle = 0$) by successive applications
of the creation operator $\ap$. Its basis states
\begin{equation}
  |n\rangle = \left(\prod_{i=1}^n F(i)\right)^{-1/2} (\ap)^n |0\rangle, \qquad n=0, 1, 2,
  \ldots,  \label{eq:basis}
\end{equation}
where we set $\prod_{i=1}^0 \equiv 1$, satisfy the relations $N |n\rangle = n |n\rangle$,
$\ap |n\rangle = \sqrt{F(n+1)} |n+1\rangle$, and $a |n\rangle = \sqrt{F(n)}
|n-1\rangle$.\par
%
%
{}For $G(N) = I$, $F(N) = N$ and the algebra \AG\ reduces to the standard (bosonic)
oscillator algebra ${\cal A}(I)$, for which the creation and annihilation operators may be
written as $\ap = (x - {\rm i}P)/\sqrt{2}$, $a = (x + {\rm i}P)/\sqrt{2}$, where $P$
denotes the momentum operator ($P = - {\rm i} d/dx$).\par
%
%
A $\Z_{\lambda}$-grading structure can be imposed on $\cal F$ by introducing a grading
operator
\begin{equation}
  T = e^{2\pi{\rm i}N/\lambda}, \qquad \lambda \in \{2, 3, 4, \ldots\},  \label{eq:T} 
\end{equation}
which is such that
\begin{equation}
  T^{\dagger} = T^{-1}, \qquad T^{\lambda} = I.  \label{eq:T-prop1}
\end{equation}
It has $\lambda$ distinct eigenvalues $q^{\mu}$, $\mu = 0$, 1,~\ldots, $\lambda-1$,
with corresponding eigenspaces ${\cal F}_{\mu} \equiv \{ |k\lambda + \mu \rangle \mid
k=0, 1, 2, \ldots\}$ such that ${\cal F} = \sum_{\mu=0}^{\lambda-1} \oplus {\cal
F}_{\mu}$. Here $q$ denotes a $\lambda$-th root of unity, $q \equiv \exp(2\pi {\rm
i}/\lambda)$, and $|n\rangle = |k\lambda + \mu\rangle$ are the basis
states~(\ref{eq:basis}). From~(\ref{eq:GDOA}), it results that $T$ satisfies the
relations
\begin{equation}
  [N, T] = 0, \qquad \ap T = e^{-2\pi{\rm i}/\lambda} T \ap, \qquad a T =
  e^{2\pi{\rm i}/\lambda} T a,  \label{eq:T-prop2} 
\end{equation}
expressing the fact that $N$ preserves the grade, while $\ap$ (resp.\ $a$) increases
(resp.\ decreases) it by one unit.\par
%
%
The operators
\begin{equation}
  P_{\mu} = \frac{1}{\lambda} \sum_{\nu=0}^{\lambda-1} e^{-2\pi{\rm
  i}\mu\nu/\lambda} T^{\nu}, \qquad \mu=0, 1, \ldots, \lambda-1,  \label{eq:P}
\end{equation}
project on the various subspaces ${\cal F}_{\mu}$, $\mu=0$, 1,~\ldots, $\lambda-1$,
and therefore satisfy the relations
\begin{equation}
  P_{\mu}^{\dagger} = P_{\mu}, \qquad P_{\mu} P_{\nu} = \delta_{\mu,\nu} P_{\mu},
  \qquad \sum_{\mu=0}^{\lambda-1} P_{\mu} = I  \label{eq:P-prop1}
\end{equation}
in $\cal F$. As a consequence of~(\ref{eq:T-prop2}), they also fulfil the relations
\begin{equation}
  [N, P_{\mu}] = 0, \qquad \ap P_{\mu} = P_{\mu+1} \ap, \qquad a P_{\mu} =
  P_{\mu-1} a,  \label{eq:P-prop2}
\end{equation}
where we use the convention 
\begin{equation}
  P_{\mu'} = P_{\mu} \qquad \mbox{\rm if\ } \mu' - \mu = 0 \mod \lambda.
  \label{eq:P-ext}
\end{equation}
\par
%
%
As a special case of GDOA with a built-in $\Z_{\lambda}$-grading structure, we may
consider the GDOA \Al\ associated with a $C_{\lambda}$-extended oscillator algebra
${\cal A}^{(\lambda)}_{\alpha_0 \alpha_1 \ldots \alpha_{\lambda-2}}$, where the
cyclic group $C_{\lambda} = \Z_{\lambda}$ is generated by $T$, i.e., $C_{\lambda} =
\{T, T^2, \ldots, T^{\lambda-1}, T^{\lambda} = I\}$ \cite{cq98, cq00}.\footnote{In
${\cal A}^{(\lambda)}_{\alpha_0 \alpha_1 \ldots \alpha_{\lambda-2}}$, $T$ is
considered as an operator independent of the remaining ones, so that
Eqs.~(\ref{eq:T-prop1}) and (\ref{eq:T-prop2}) (or alternatively (\ref{eq:P-prop1})
and (\ref{eq:P-prop2})) have to be postulated in addition to (\ref{eq:GDOA}) and
(\ref{eq:G}). The GDOA \Al\ corresponds to the realization (\ref{eq:T}) of $T$.}  The
GDOA \Al\ corresponds to the choice
\begin{equation}
  G(N) = I + \sum_{\mu=1}^{\lambda-1} \kappa_{\mu} T^{\mu} = I +
  \sum_{\mu=0}^{\lambda-1} \alpha_{\mu} P_{\mu}  \label{eq:G}
\end{equation}
in Eq.~(\ref{eq:GDOA}). Here $\kappa_{\mu}$, $\mu=1$, 2,~\ldots, $\lambda-1$, are
some complex parameters subject to the conditions $\kappa_{\mu}^* =
\kappa_{\lambda - \mu}$, while $\alpha_{\mu}$, $\mu=0$, 1,~\ldots, $\lambda-1$, are
some real ones constrained by $\sum_{\mu=0}^{\lambda-1} \alpha_{\mu} = 0$. Hence
the algebra depends upon $\lambda-1$ independent, real parameters, which may be
chosen as
$\alpha_0$, $\alpha_1$,~\ldots, $\alpha_{\lambda-2}$.\par
%
%
The corresponding structure function can be expressed as 
\begin{equation}
  F(N) = N + \sum_{\mu=0}^{\lambda-1} \beta_{\mu} P_{\mu}, \qquad \beta_0 \equiv
  0, \qquad \beta_{\mu} \equiv \sum_{\nu=0}^{\mu-1} \alpha_{\nu}, \qquad \mu=1, 2,
  \ldots, \lambda-1.  \label{eq:F}
\end{equation}
The existence of a Fock-space representation is guaranteed by the $\lambda-1$
constraints on the parameters $\sum_{\nu=0}^{\mu} \alpha_{\nu} > - \mu - 1$,
$\mu=0$, 1,~\ldots,
$\lambda-2$, ensuring that $F(\mu) = \mu + \beta_{\mu} > 0$ for $\mu=1$, 2,~\ldots,
$\lambda-1$.\par
%
%
The standard oscillator algebra ${\cal A}(I)$ is recovered in the limit $\alpha_{\mu} \to
0$, $\mu = 0$, 1, \ldots,~$\lambda-1$.\par
%
%
\section{\boldmath Order-$p$ Parasupersymmetric Quantum Mechanics}

PSSQM of order $p$ is described in terms of parasupercharge operators $Q$, $\Qp$,
and of a parasupersymmetric Hamiltonian $\cal H$ satisfying the relations
\begin{eqnarray}
  Q^{p+1} & = & 0 \qquad (\mbox{\rm with\ } Q^p \ne 0),  \label{eq:PSSQM1} \\[0pt]
  [{\cal H}, Q] & = & 0,  \label{eq:PSSQM2}
\end{eqnarray}
and
\begin{equation}
  Q^p \Qp + Q^{p-1} \Qp Q + \cdots + Q \Qp Q^{p-1} + \Qp Q^p = 2p Q^{p-1} {\cal H}
  \label{eq:PSSQM3-1}
\end{equation}
or
\begin{equation}
  [Q, [\Qp, Q]] = 2Q {\cal H},  \label{eq:PSSQM3-2}
\end{equation}
according to whether one considers the RSK approach~\cite{rubakov} or the BD
one~\cite{beckers90}. They also obey the Hermitian conjugated relations with
$(\Qp)^{\dagger} = Q$ and ${\cal H}^{\dagger} = {\cal H}$.\par
%
%
To start with, it is easy to get a $(p+1) \times (p+1)$-matrix realization of RSK PSSQM of
the type
\begin{equation}
  Q = \sqrt{2} \sum_{i=1}^p f_i(N+1) a e_{i+1,i}, \qquad \Qp = \sqrt{2} \sum_{i=1}^p
  f_i(N) \ap e_{i,i+1}, \qquad {\cal H} = \sum_{i=1}^{p+1} H_i e_{i,i}, 
  \label{eq:PSSQM-matrix1} 
\end{equation}
in terms of the generators $N$, $\ap$, $a$ of a GDOA \AG. Here $f_i(N)$, $i=1$,
2,~\ldots, $p$, are assumed to be some real functions of $N$ defined on the set $\{1,
2, 3, \ldots\}$, and $H_i$, $i=1$, 2,~\ldots, $p+1$, some $N$-dependent Hermitian
operators, while $e_{i,j}$ denotes the $(p+1)$-dimensional matrix with entry 1 at the
intersection of row $i$ and column $j$ and zeroes everywhere else. The
ansatz~(\ref{eq:PSSQM-matrix1}) indeed automatically satisfies Eq.~(\ref{eq:PSSQM1}).
Inserting it into the remaining two equations (\ref{eq:PSSQM2}) and (\ref{eq:PSSQM3-1})
leads to an explicit form for the Hamiltonians $H_i$, 
\begin{equation}
  H_i = \frac{1}{p} \sum_{j=1}^p f_j^2(N+i-j) F(N+i-j), \qquad i=1, 2, \ldots, p+1,
  \label{eq:PSSQM-Hi-1}
\end{equation}
in terms of the $p$ arbitrary functions $f_i(N)$, $i=1$, 2,~\ldots, $p$, and of the
structure function $F(N)$, solution of Eqs.~(\ref{eq:G-F}) and~(\ref{eq:F-cond}).\par
%
%
It is worth mentioning that for $G(N) = I$ (corresponding to the standard oscillator
algebra) and $f_i(N) = 1$, $i=1$, 2,~\ldots, $p$, the parasupersymmetric Hamiltonian
$\cal H$, as defined in (\ref{eq:PSSQM-matrix1}) and (\ref{eq:PSSQM-Hi-1}), reduces to
the RSK parasupersymmetric oscillator one, ${\cal H}^{\rm RSK}_{\rm osc} = \frac{1}{2}
(P^2 + x^2) \I - \diag\left(\frac{p}{2}, \frac{p}{2} - 1, \ldots, - \frac{p}{2} + 1, -
\frac{p}{2}\right)$, where \I\ denotes the $(p+1) \times (p+1)$ unit matrix. The
corresponding charges $Q = {\rm i}(P - {\rm i} x) \sum_{i=1}^p e_{i+1,i}$, $\Qp = - {\rm
i}(P + {\rm i} x) \sum_{i=1}^p e_{i,i+1}$ only differ from the RSK ones by an irrelevant
phase factor.\par
%
%
The $(p+1) \times (p+1)$-matrix realization (\ref{eq:PSSQM-matrix1}),
(\ref{eq:PSSQM-Hi-1}) of RSK PSSQM can be diagonalized through a unitary
transformation $U_1 = \sum_{i,j=1}^{p+1} P_{i-j} e_{i,j}$, expressed in terms of the
projection operators $P_{\mu}$ defined in (\ref{eq:P}) and (\ref{eq:P-ext}) for $\lambda
= p+1$. The results read
\begin{eqnarray}
  Q' & \equiv & U_1 Q U_1^{\dagger} = \diag(Q_0, Q_1, \ldots, Q_p), \nonumber \\
  Q^{\prime\dagger} & \equiv & U_1 \Qp U_1^{\dagger} = \diag(\Qp_0, \Qp_1, \ldots,
       \Qp_p),  \\
  {\cal H}' & \equiv & U_1 {\cal H} U_1^{\dagger} = \diag({\cal H}_0, {\cal H}_1, \ldots,
       {\cal H}_p),  \nonumber  
\end{eqnarray}
where
\begin{eqnarray}
  Q_{\mu} & = & \sqrt{2} \sum_{i=1}^p f_i(N+1) a P_{\mu+p+2-i}, \qquad 
        \Qp_{\mu} = \sqrt{2} \sum_{i=1}^p f_i(N) \ap P_{\mu+p+1-i},
        \label{eq:PSSQM-Qmu} \\
  {\cal H}_{\mu} & = & \sum_{i=1}^{p+1} H_i P_{\mu+p+2-i},  \label{eq:PSSQM-Hmu}
\end{eqnarray}
for $\mu=0$, 1,~\ldots, $p$. Each of the $p+1$ sets of operators $\{Q_{\mu},
\Qp_{\mu}, {\cal H}_{\mu}\}$ satisfies the RSK PSSQM algebra (\ref{eq:PSSQM1}) --
(\ref{eq:PSSQM3-1}) and is written in terms of a single bosonic degree of freedom through
the operators $N$, $\ap$, $a$ of \AG. We have therefore proved that RSK PSSQM is fully
reducible and, in addition, we have obtained a minimal bosonization thereof.\par
%
%
The eigenvalues ${\cal E}^{(\mu)}_n$ of the bosonized parasupersymmetric Hamiltonian
${\cal H}_{\mu}$, defined in (\ref{eq:PSSQM-Hmu}), can be written as
\begin{eqnarray}
  {\cal E}^{(\mu)}_{k(p+1)+\nu} & = & \frac{1}{p} \sum_{i=1}^p f_i^2[k(p+1) + \mu -
        i + 1] F[k(p+1) + \mu - i + 1] \nonumber \\
  && \mbox{\rm if\ } \nu=0, 1, \ldots, \mu,  \label{eq:PSSQM-E1-1} \\
  {\cal E}^{(\mu)}_{k(p+1)+\nu} & = & \frac{1}{p} \sum_{i=1}^p f_i^2[(k+1)(p+1) +
        \mu - i + 1] F[(k+1)(p+1) + \mu - i + 1] \nonumber \\
  && \mbox{\rm if\ } \nu=\mu+1, \mu+2, \ldots, p,  \label{eq:PSSQM-E1-2}  
\end{eqnarray}
and therefore correspond to nonlinear spectra. At this stage, we may remark that since
the sign of the structure function for negative integers is not fixed by
condition~(\ref{eq:F-cond}), ${\cal E}^{(\mu)}_{\nu}$ may be positive, null, or negative
for $\nu=0$, 1,~\ldots, $\mu$ and $\mu=0$, 1,~\ldots, $p-2$. We therefore recover
here a well-known unsatisfactory feature of RSK PSSQM~\cite{rubakov}.\par
%
%
The results obtained so far may be illustrated by considering the GDOA ${\cal
A}^{(p+1)}(G(N))$, defined in Sec.~2. This amounts to inserting the structure function
(\ref{eq:F}) for $\lambda=p+1$ into Eqs.~(\ref{eq:PSSQM-Hi-1}),
(\ref{eq:PSSQM-E1-1}), and~(\ref{eq:PSSQM-E1-2}). If $f_i^2(n)$, $i=1$, 2,~\ldots,
$p$, are chosen to be increasing functions of $n$, the ground state of the spectrum
(\ref{eq:PSSQM-E1-1}), (\ref{eq:PSSQM-E1-2}) corresponds to $k=0$, $\nu=0$,
1,~\ldots, $\mu$, and is $(\mu+1)$-fold degenerate, while all the excited states are
$(p+1)$-fold degenerate. Since $\mu$ may take any value in the set $\{0, 1, \ldots,
p\}$, the ground-state degeneracy may accordingly vary between 1 and $p+1$. As
hereabove noted, it results from the constraints on the algebra parameters that the
ground-state energy
\begin{eqnarray}
  {\cal E}^{(\mu)}_0 & = & {\cal E}^{(\mu)}_1 = \cdots = {\cal E}^{(\mu)}_{\mu}
        = \frac{1}{p} \Biggl[\sum_{i=1}^{\mu} f_i^2(\mu-i+1) (\mu-i+1 +
        \beta_{\mu-i+1}) \nonumber \\
  && \mbox{} + \sum_{i=\mu+2}^p f_i^2(\mu-i+1) (\mu-i+1 +
        \beta_{\mu+p+2-i})\Biggr]  
\end{eqnarray}
may have any sign except for $\mu=p-1$ or $\mu=p$, for which it is positive. It can be
checked that $Q_0$ and $\Qp_0$ have a vanishing action on $|0\rangle$, but that this
is not true for $Q_{\mu}$, $\Qp_{\mu}$ on $|0\rangle$, $|1\rangle$, \ldots,
$|\mu\rangle$, when $\mu=1$, 2,~\ldots, or $p$. We conclude that unbroken (resp.\
broken) PSSQM corresponds to $\mu=0$ (resp.\ $\mu=1$, 2,~\ldots, or $p$).\par
%
%
In the special case where $f_i(N)=1$, $i=1$, 2,~\ldots, $p$, the bosonized
parasupersymmetric Hamiltonian (\ref{eq:PSSQM-Hmu}) associated with ${\cal
A}^{(p+1)}(G(N))$ reduces to the operator (4.34) found in Ref.~\cite{cq00} and giving
rise to a linear spectrum, while the bosonized parasupercharges (\ref{eq:PSSQM-Qmu})
become the corresponding charges (up to an interchange of $Q_{\mu}$ and
$\Qp_{\mu}$, which leaves the PSSQM algebra invariant).\par
%
%
Let us now consider the case of BD PSSQM, defined by Eqs.~(\ref{eq:PSSQM1}),
(\ref{eq:PSSQM2}), and (\ref{eq:PSSQM3-2}), and choose an ansatz similar
to~(\ref{eq:PSSQM-matrix1}), except for a convenient renormalization of the functions
$f_i(N)$,
\begin{eqnarray}
  Q & = & \sum_{i=1}^p [i(p-i+1)]^{1/2} f_i(N+1) a e_{i+1,i}, \quad \Qp = 
        \sum_{i=1}^p [i(p-i+1)]^{1/2} f_i(N) \ap e_{i,i+1}, \nonumber \\
  {\cal H} & = & \sum_{i=1}^{p+1} H_i e_{i,i}.  \label{eq:PSSQM-matrix2} 
\end{eqnarray}
It is easy to show that Eqs.~(\ref{eq:PSSQM2}) and (\ref{eq:PSSQM3-2}) are satisfied
provided
\begin{equation}
  H_i = f_1^2(N+i-1) F(N+i-1), \qquad i=1, 2, \ldots, p+1, \label{eq:PSSQM-Hi-2}
\end{equation}
and, in addition, the functions $f_i(N)$ fulfil the conditions
\begin{equation}
  f_i(N) \sqrt{F(N)} = \epsilon_i(N) f_1(N+i-1) \sqrt{F(N+i-1)}, \qquad i = 2, 3, \ldots, p,
  \label{eq:PSSQM-f-cond}
\end{equation}
where $\epsilon_i^2(N) = 1$. The general solution of (\ref{eq:PSSQM-f-cond}) can be
written as
\begin{equation}
  f_i(N) = \epsilon_i(N) g(N+i-1) \left(\prod_{\stackrel{j=1}{j\ne i}}^p
  F(N+i-j)\right)^{1/2}, \qquad i=1, 2, \ldots, p, \label{eq:PSSQM-f-sol}
\end{equation}
with $\epsilon_1(N) \equiv 1$ and $g(N)$ an arbitrary real function of $N$. Hence
Eq.~(\ref{eq:PSSQM-Hi-2}) finally becomes
\begin{equation}
  H_i = g^2(N+i-1) \prod_{j=1}^p F(N+i-j), \qquad i=1, 2, \ldots, p+1.
  \label{eq:PSSQM-Hi-2bis}
\end{equation}
\par
%
%
As in the RSK case, the $(p+1)$-dimensional matrix realization of BD PSSQM, given in
(\ref{eq:PSSQM-matrix2}), (\ref{eq:PSSQM-f-sol}), and (\ref{eq:PSSQM-Hi-2bis}), can be
diagonalized through the unitary transformation $U_1$. The bosonized parasupercharges
and parasupersymmetric Hamiltonian are still given by Eqs.~(\ref{eq:PSSQM-Qmu}) and
(\ref{eq:PSSQM-Hmu}) (up to the renormalization $\sqrt{2} f_i(N) \to [i(p-i+1)]^{1/2}
f_i(N)$), where $f_i(N)$ and $H_i$ are now expressed in the form
(\ref{eq:PSSQM-f-sol}) and (\ref{eq:PSSQM-Hi-2bis}), respectively.\par
%
%
The eigenvalues ${\cal E}^{(\mu)}_n$ of ${\cal H}_{\mu}$ can be written as
\begin{eqnarray}
  {\cal E}^{(\mu)}_{k(p+1)+\nu} & = & g^2[k(p+1) + \mu] \prod_{i=1}^p  F[k(p+1) +
        \mu - i + 1] \nonumber \\ 
  && \mbox{\rm if\ } \nu=0, 1, \ldots, \mu, \label{eq:PSSQM-E2-1} \\
  {\cal E}^{(\mu)}_{k(p+1)+\nu} & = & g^2[(k+1)(p+1) + \mu] \prod_{i=1}^p  
        F[(k+1)(p+1) + \mu - i + 1] \nonumber \\
  && \mbox{\rm if\ } \nu=\mu+1, \mu+2, \ldots, p. 
\end{eqnarray}
Here we may remark that contrary to what happens in the RSK case, no eigenvalue can be
negative because every time a structure function with negative argument appears in
(\ref{eq:PSSQM-E2-1}), it is multiplied by $F(0)$, which is vanishing according
to~(\ref{eq:F-cond}). We therefore recover a property of BD
PSSQM~\cite{beckers90}.\par
%
%
{}For the GDOA ${\cal A}^{(p+1)}(G(N))$ and an increasing function $g^2(n)$, the
properties of the spectrum are similar to those obtained hereabove in the RSK case,
except that now the ground-state energy ${\cal E}^{(\mu)}_0 = {\cal E}^{(\mu)}_1 =
\cdots = {\cal E}^{(\mu)}_{\mu}$ vanishes for $\mu=0$, 1,~\ldots, $p-1$ and is equal
to $g^2(p) \prod_{i=1}^p (p-i+1 + \beta_{p-i+1}) > 0$ for $\mu=p$. It can be checked
that the former (resp.\ latter) case corresponds to unbroken (resp.\ broken) PSSQM.
This agrees with the detailed study of $p=2$ parasuperspectra carried out in
Ref.~\cite{beckers90}.\par
%
%
Had we chosen $f_i(N) = 1$, $i=1$, 2,~\ldots, $p$, from the very beginning in
Eq.~(\ref{eq:PSSQM-matrix2}), the conditions (\ref{eq:PSSQM-f-cond}) would lead to
some constraints on the structure function. For $p=2$, for instance, we would get a single constraint $F(N)
= F(N+1)$. For the GDOA ${\cal A}^{(p+1)}(G(N))$, it gives rise to the conditions
$\alpha_{\mu} = -1$, $\mu=0$, 1, 2, which are incompatible with the existence of a Fock
space representation (see also Ref.~\cite{cq98}). We conclude that for the proof of the
reducibility of BD PSSQM and its resultant bosonization, the presence of nontrivial
functions $f_i(N)$ in (\ref{eq:PSSQM-matrix2}) plays a crucial role.\par
%
%
It should be noted that the BD parasupersymmetric oscillator
Hamiltonian~\cite{beckers90} cannot be seen as a special case of the matrix
realization~(\ref{eq:PSSQM-matrix2}) when $G(N) \to I$. There actually exists another
$(p+1) \times (p+1)$-matrix realization of BD PSSQM in terms of \AG\ generators that
has this property: instead of having $a$ everywhere below the principal diagonal in $Q$,
it has $\ap$ and $a$ appearing there in turn. Such a realization is however not reducible
along the lines of the present work.\par
%
%
\section{\boldmath Order-$p$ Orthosupersymmetric Quantum Mechanics}

OSSQM of order $p$ is formulated in terms of an orthosupersymmetric Hamiltonian $\cal
H$ and $p$ pairs of orthosupercharge operators $Q_i$, $\Qp_i$, $i=1$, 2,~\ldots, $p$,
satisfying the relations
\begin{eqnarray}
  Q_i Q_j & = & 0, \label{eq:OSSQM1} \\[0pt]
  [{\cal H}, Q_i] & = & 0, \\
  Q_i \Qp_j + \delta_{i,j} \sum_{k=1}^p \Qp_k Q_k & = & 2 \delta_{i,j} {\cal H},
       \label{eq:OSSQM3}
\end{eqnarray}
and their Hermitian conjugates with $(\Qp_i)^{\dagger} = Q_i$ and ${\cal H}^{\dagger}
= {\cal H}$, where $i$, $j$ run over 1, 2,~\ldots, $p$~\cite{khare}.\par
%
%
A $(p+1) \times (p+1)$-matrix realization in terms of \AG\ generators can be found by
setting
\begin{equation}
  Q_i = \sqrt{2} f_i(N+i) a^i e_{1,i+1}, \qquad \Qp_i = \sqrt{2} f_i(N) (\ap)^i e_{i+1,1},
  \qquad {\cal H} = \sum_{i=1}^{p+1} H_i e_{i,i},  \label{eq:OSSQM-matrix} 
\end{equation}
where $f_i(N)$, $i=1$, 2,~\ldots, $p$, are some real functions of $N$, defined on the set
$\{i, i+1, \ldots\}$, and $H_i$, $i=1$, 2,~\ldots, $p+1$, are some $N$-dependent
Hermitian operators. One indeed finds that Eqs.~(\ref{eq:OSSQM1}) --
(\ref{eq:OSSQM3}) are satisfied provided
\begin{equation}
  H_i = f_p^2(N+p+1-i) \prod_{j=1}^p F(N+j+1-i), \qquad i=1, 2, \ldots, p+1,
  \label{eq:OSSQM-Hi}
\end{equation}
and 
\begin{equation}
  f_i(N) = \epsilon_i(N) f_p(N+p-i) \left(\prod_{j=1}^{p-i} F(N+j)\right)^{1/2}, \qquad
  i= 1, 2, \ldots, p, \label{eq:OSSQM-fi}
\end{equation}
where $\epsilon_i^2(N) = 1$.\par
%
%
The matrices (\ref{eq:OSSQM-matrix}) can be reduced by means of the unitary
transformation $U_2 = \sum_{i,j=1}^{p+1} P_{i+j-1} e_{i,j}$, where the $P_{\mu}$'s
are again the projection operators (\ref{eq:P}), (\ref{eq:P-ext}) for $\lambda = p+1$. As
a consequence, we obtain $p+1$ sets of bosonized orthosupercharges and
orthosupersymmetric Hamiltonian
\begin{eqnarray}
  Q_{i\mu} & = & \sqrt{2} f_i(N+i) a^i P_{\mu+i+1}, \qquad \Qp_{i\mu} = \sqrt{2} f_i(N)
         (\ap)^i P_{\mu+1}, \label{eq:OSSQM-Qmu} \\
  {\cal H}_{\mu} & = & \sum_{i=1}^{p+1} H_i P_{\mu+i}, \label{eq:OSSQM-Hmu} 
\end{eqnarray}
where $\mu=0$, 1,~\ldots, $p$, and $H_i$, $f_i(N)$ are given in
Eqs.~(\ref{eq:OSSQM-Hi}), (\ref{eq:OSSQM-fi}), respectively.\par
%
%
The eigenvalues ${\cal E}^{(\mu)}_n$ of ${\cal H}_{\mu}$, defined in
(\ref{eq:OSSQM-Hmu}), can be written as
\begin{eqnarray}
  {\cal E}^{(\mu)}_{k(p+1)+\nu} & = & f_p^2[k(p+1)+\mu] \prod_{i=1}^p
         F[k(p+1)+\mu-p+i] \nonumber \\
  && \mbox{\rm if\ } \nu=0, 1, \ldots, \mu, \\
  {\cal E}^{(\mu)}_{k(p+1)+\nu} & = & f_p^2[(k+1)(p+1)+\mu] \prod_{i=1}^p
         F[(k+1)(p+1)+\mu-p+i] \nonumber \\ 
  &&\mbox{\rm if\ } \nu=\mu+1, \mu+2, \ldots, p,
\end{eqnarray}
and therefore correspond to nonlinear spectra again. As in the case of BD PSSQM, any
structure function with negative argument is multiplied by $F(0)$, hence ${\cal
E}^{(\mu)}_{\nu}$ is always nonnegative.\par
%
%
When we choose the GDOA ${\cal A}^{(p+1)}(G(N))$ and an increasing function
$f_p^2(n)$, we find that for $\mu=0$, 1,~\ldots, $p-1$, the orthosupersymmetry is
unbroken and the ground state, corresponding to $k=0$, $\nu=0$, 1,~\ldots, $\mu$,
has a vanishing energy and a degeneracy $\mu+1$. On the contrary, for $\mu=p$, the
orthosupersymmetry is broken and the $(p+1)$-fold degenerate ground state has a
positive energy, given by ${\cal E}^{(p)}_0 = {\cal E}^{(p)}_1 = \cdots = {\cal
E}^{(p)}_p = f_p^2(p) \prod_{i=1}^p (i + \beta_i)$. In both cases, all the excited
states have a degeneracy $p+1$. Such results are in accordance with those of Khare
{\em et al.}~\cite{khare}.\par
%
%
As for BD PSSQM, the presence of nontrivial functions $f_i(N)$ in
(\ref{eq:OSSQM-matrix}) is essential to prove the reducibility and bosonization of OSSQM
in full generality. Setting $f_i(N) = 1$, $i=1$, 2,~\ldots, $p$, in (\ref{eq:OSSQM-fi})
would indeed lead to some restrictions on the structure function.\par
%
%
The bosonized operators (\ref{eq:OSSQM-Qmu}), (\ref{eq:OSSQM-Hmu}) differ from
those previously obtained for $p=2$ and linear spectra, which are linear in the creation
and annihilation operators~\cite{cq00}. The latter may actually be generalized to
nonlinear spectra by introducing some $N$-dependent coefficients and they may be
derived by reducing some appropriate $3 \times 3$-matrix realization of $p=2$ OSSQM.
A straightforward extension to higher values of $p$ however proves impossible. To get
round this difficulty, it is necessary to consider powers of the creation and annihilation
operators, as we did in (\ref{eq:OSSQM-matrix}) and (\ref{eq:OSSQM-Qmu}). As a
consequence, the orthosupersymmetric oscillator Hamiltonian~\cite{khare} cannot be
retrieved in the limit $G(N) \to I$.\par
%
%
\section{Conclusion}

In the present work, we did prove that both versions of order-$p$ PSSQM, as well as
order-$p$ OSSQM, are fully reducible when we realize them in terms of GDOA generators
and impose a $\Z_{p+1}$-grading on the Fock space. As a consequence, we did get a
{\em minimal\/} bosonization of such SSQM variants in terms of a single bosonic degree
of freedom. Similar results have been obtained elsewhere in the case of
PsSSQM~\cite{cq02}. It is worth stressing that everything remains true in the limit $G(N)
\to I$, corresponding to the standard bosonic oscillator algebra.\par
%
%
In the cases of BD PSSQM anf of OSSQM, our results were obtained at the expense of
considering Hamiltonians that in the limit $G(N) \to I$, contain powers of $P^2$ and
would therefore perhaps more appropriately be called `quasi-Hamiltonians'. In addition,
the OSSQM charges contain powers of $P$ in the same limit. Since such features are
characteristic of higher-derivative SSQM~\cite{andrianov} and $\cal N$-fold
SSQM~\cite{aoyama}, the existence of connections with these theories is an interesting
open question, which we hope to discuss in a near future.\par
%
%

\end{document}